\documentclass[aps,prl,twocolumn,eqnum,showpacs,amsmath,nofootinbib]{revtex4}%

\usepackage[dvips]{color,graphicx}
\usepackage{amsfonts,amssymb,theorem,mathrsfs,times}
\newcommand{\ma}[1]{\mbox{$\mathcal{#1}$}}
\newcommand{\qed}{\hbox{\rule[-2pt]{6pt}{6pt}}}
\newcommand{\D}{{\rm d}}

{\theorembodyfont{\upshape}
\newtheorem{Prop}{Proposition}}
{\theorembodyfont{\upshape}
\newtheorem{The}{Theorem}}
{\theorembodyfont{\upshape}
\newtheorem{lm}{Lemma}}
{\theorembodyfont{\upshape}
}
{\theorembodyfont{\upshape}
}
{\theorembodyfont{\upshape}
}

\newcommand{\dalm}{\kern1pt\vbox{\hrule height 0.9pt\hbox{\vrule width
0.9pt\hskip 2.5pt\vbox{\vskip 5.5pt}\hskip 3pt\vrule width 0.3pt}\hrule height
0.3pt}\kern1pt}

\begin{document}

\title{
Universal slow fall-off to the unique AdS infinity in Einstein-Gauss-Bonnet gravity
}

\author{Hideki Maeda}
\email{hideki@cecs.cl}


\address{ 
Centro de Estudios Cient\'{\i}ficos (CECS), Arturo Prat 514, Valdivia, Chile
}

\date{\today}

\begin{abstract} 
In this paper, the following two propositions are proven under the dominant energy condition for the matter field in the higher-dimensional spherically symmetric spacetime in Einstein-Gauss-Bonnet gravity in the presence of a cosmological constant $\Lambda$.
First, for $\Lambda\le 0$ and $\alpha \ge 0$ without a fine-tuning to give a unique anti-de~Sitter (AdS) vacuum, where $\alpha$ is the Gauss-Bonnet coupling constant, vanishing generalized Misner-Sharp mass is equivalent to the maximally symmetric spacetime.
Under the fine-tuning, it is equivalent to the vacuum class I spacetime.
Second, under the fine-tuning with $\alpha>0$, the asymptotically AdS spacetime in the higher-dimensional Henneaux-Teitelboim sense is only a special class of the vacuum class I spacetime.
This means the universal slow fall-off to the unique AdS infinity in the presence of physically reasonable matter.
\end{abstract}

\pacs{
04.20.Cv, 
04.20.Ha,
04.50.-h. 
} 


\maketitle

{\it 1. Introduction:}
Gravitation physics in higher dimensions is a prevalent subject of current research motivated mainly by string theory. 
In this context, it is well known that the most natural extension of general relativity in higher dimensions as a theory of quasi-linear second-order partial differential equations is not general relativity itself but Lovelock gravity~\cite{lovelock}.

The Lovelock Lagrangian comprises the dimensionally extended Euler densities.
In $n$ dimensions, the first [$n/2$] curvature terms appear in the field equations, where [$x$] denotes the integer part of $x$. 
In the even-dimensional case, however, the last ($(n/2)$-th) term becomes a topological invariant and does not contribute to the field equations. 
The Lovelock tensor ${\ma G}_{\mu \nu }$ derived from the Lovelock Lagrangian has the following properties:
(1) ${\ma G}_{\mu \nu }$ is symmetric,
(2) ${\ma G}_{\mu \nu }$ contains up to the second derivative of the metric,
(3) $\nabla _\nu {\ma G}^{\mu\nu} \equiv 0$, and 
(4) ${\ma G}_{\mu \nu }$ is linear in the second derivative of the metric.

Lovelock gravity, as well as general relativity, is a gauge theory for the (local) Lorentz group obviously but not for the Poincar{\'e} group in general, which is the standard symmetry group in particle physics~\cite{zanelli2005}.
From the gauge-principle viewpoint, the gravitation theory is expected to be a gauge theory for the Poincar{\'e} group or some group which contains the Lorentz group and the symmetry group analogous to translations in a flat spacetime.
Miraculously, under the fine-tuning between the coupling constants, Lovelock gravity can be a gauge theory for the Poincar{\'e}, de~Sitter (dS), or anti-de~Sitter (AdS) group.
The last two groups are the smallest nontrivial choices of such required groups containing the translation symmetry group on a pseudosphere.
Unfortunately, this miracle happens only in {\it odd} dimensions, nevertheless this so-called Chern-Simons gravity has been of particular interest as an aesthetic way to the unified theory~\cite{zanelli2005}.

On the other hand, the second-order Lovelock theory so-called Einstein-Gauss-Bonnet gravity has been intensively investigated because the renormalizable Gauss-Bonnet term appears in the low-energy limit of the heterotic string theory~\cite{Gross}.
The history of the black-hole physics in Einstein-Gauss-Bonnet gravity began from the well-known Boulware-Deser-Wheeler solution corresponding to the Schwarzschild-Tangherlini solution in general relativity~\cite{GB_BH}.
This solution has been generalized~\cite{GBBH-solution} and occupied the central position in the research of the Gauss-Bonnet black holes. 
Based on this generalized Boulware-Deser-Wheeler solution, the effects of the Gauss-Bonnet term on the stability~\cite{GBBH-stability} and the black-hole thermodynamics~\cite{GBBH-thermodynamics} have been investigated.
(We refer~\cite{lovelockBH} for the recent review.)
Intriguingly, the solutions in Einstein-Gauss-Bonnet gravity are classified into two branches, one of which admits the general relativistic limit, while the other does not.
As a result, the theory generically admits two distinct (A)dS vacua stemming from its quadratic nature. 

Recently, asymptotically AdS black holes with a scalar hair have attracted much attention, which was first discovered numerically as a counterexample of the black-hole no-hair conjecture~\cite{scalar-hair}.
In particular, theories of AdS gravity coupled to a scalar field with mass at or slightly above the Breitenlohner-Freedman bound~\cite{bf} are called {\it designer gravity}~\cite{designer}.
Designer gravity admits a large class of asymptotically AdS spacetimes with slower fall-off conditions than the standard ones, of which boundary conditions are defined by an essentially arbitrary function. 
In the context of the AdS/CFT correspondence~\cite{ads/cft}, asymptotically AdS black-hole solutions in designer gravity have been applied to the study of the cosmic censorship conjecture~\cite{designer-cch} or big-bang singularities~\cite{designer-cosmology}.

In the present paper, we show that such remarkable slow fall-off to the AdS infinity is universal in spherically symmetric spacetimes containing {\it any} matter satisfying the dominant energy condition in Einstein-Gauss-Bonnet gravity with a fine-tuning of the coupling constants to give a unique AdS vacuum, where the theory becomes Chern-Simons gravity in five dimensions. 
We adopt the units in which only the $n$-dimensional gravitational constant $G_n$ is retained.

{\it 2. Preliminaries:}
The field equation of Einstein-Gauss-Bonnet gravity in the $n (\geq 5)$-dimensional spacetime is
\begin{align}
&{G^\mu}_{\nu} +\alpha {H}^\mu_{~~\nu} 
+\Lambda \delta^\mu_{~~\nu}= 
\kappa_n^2 {T}^\mu_{~~\nu}, \label{beq} \\
&{G}_{\mu\nu}:= R_{\mu\nu}-{1\over 2}g_{\mu\nu}R,\\
&{H}_{\mu\nu}:= 2(RR_{\mu\nu}-2R_{\mu\alpha}
R^\alpha_{~\nu}-2R^{\alpha\beta}R_{\mu\alpha\nu\beta}+R_{\mu}^{~\alpha\beta\gamma}R_{\nu\alpha\beta\gamma})
 \nonumber \\
&~~~~~~-{1\over 2}g_{\mu\nu}(R^2-4R_{\mu\nu}R^{\mu\nu}
+R_{\mu\nu\rho\sigma}R^{\mu\nu\rho\sigma}),
\end{align}
where $\kappa_n := \sqrt{8\pi G_n}$ and $\Lambda$ is a cosmological constant.
$\alpha$ is the Gauss-Bonnet coupling constant and ${T}^\mu_{~~\nu}$ is the energy-momentum tensor for matter fields.

Suppose the $n$-dimensional spacetime 
$({\ma M}^n, g_{\mu \nu })$ to be a warped product of an 
$(n-2)$-dimensional constant curvature space $(K^{n-2}, \gamma _{ij})$ with its sectional curvature $k = \pm 1, 0$
and a two-dimensional orbit spacetime $(M^2, g_{ab})$ under 
the isometries of $(K^{n-2}, \gamma _{ij})$. 
We assume that $K^{n-2}$ is compact. 

The line element in the double-null coordinates is given by
\begin{align}
\D s^2 = -2e^{-f(u,v)}\D u\D v
+r^2(u,v) \gamma_{ij}\D z^i\D z^j. \label{coords}
\end{align}  
The metric functions $e^{-f}$ and $r^2$ are non-zero and finite to avoid the coordinate singularities.
Null vectors $(\partial /\partial u)$ and $(\partial /\partial v)$ 
are taken to be future-pointing. 
The area expansions along these two radial null vectors\footnote{In~\cite{mn2008,nm2008}, they are erroneously mentioned as the expansions of the future-directed radial null geodesics.} are given as $\theta_{+}:=(n-2)r^{-1}r_{,v}$ and $\theta_{-}:=(n-2)r^{-1}r_{,u}$.
An $(n-2)$-surface with $\theta_{+}\theta_{-}>(<)0$ is called a {\it trapped (untrapped)} surface.
We fix the orientation of the untrapped surface by
$\theta _+>0$ and $\theta_-<0$, i.e., $\partial/\partial u$ and $\partial/\partial v$ are ingoing and outgoing null vectors, respectively.
The generalized Misner-Sharp mass is given by 
\begin{align}
\label{qlm2}
m &= \frac{(n-2)V_{n-2}^k}{2\kappa_n^2}r^{n-3}
\biggl[-{\tilde \Lambda}r^2+\left(k+\frac{2r^2e^{f}}{(n-2)^2}
\theta_{+}\theta_{-}\right)\nonumber \\
&~~~~~~
+{\tilde \alpha}r^{-2}\left(k+\frac{2r^2e^{f}}{(n-2)^2}\theta_{+}\theta_{-}\right)^2\biggl], 
\end{align}  
where ${\tilde \alpha} := (n-3)(n-4)\alpha$ and ${\tilde \Lambda} := 2\Lambda /[(n-1)(n-2)]$ and $V_{n-2}^k$ denotes the area of $K^{n-2}$~\cite{maeda2006b,mn2008}.
For $1+4{\tilde\alpha}{\tilde\Lambda}=0$, we have  
\begin{align}
m = \frac{(n-2)V_{n-2}^kr^{n-5}}{8{\tilde \alpha}\kappa_n^2}
\biggl[r^2+2{\tilde \alpha}\left(k+\frac{2r^2e^{f}}{(n-2)^2}\theta_{+}\theta_{-}\right)\biggl]^2, \label{qlm-sp}
\end{align}  
which is non-negative (non-positive) for $\alpha > (<)0$.

The most general $T_{\mu\nu}$ in this spacetime 
is given by
\begin{align}
T_{\mu\nu}\D x^\mu \D x^\nu =
&T_{uu}(u,v)\D u^2+2T_{uv}(u,v)\D u\D v \nonumber \\
&
+T_{vv}(u,v)\D v^2+p(u,v)r^2 \gamma_{ij}\D z^i\D z^j.
\end{align}  
The variation of $m$ is determined by the field equations as
\begin{align}
m_{,v}&=
\frac{1}{n-2}V_{n-2}^ke^fr^{n-1}(T_{uv}\theta_+-T_{vv}\theta_-), \label{m_v} \\
m_{,u}&=
\frac{1}{n-2}V_{n-2}^ke^fr^{n-1}(T_{uv}\theta_- -T_{uu}\theta_+). \label{m_u} 
\end{align}  
We assume the dominant energy condition for the matter field, which implies
\begin{align}
T_{uu} \ge 0,~~T_{vv}\ge 0,~~T_{uv}\ge 0. \label{dec}
\end{align}

For the proof of our main results, we review the generalized Birkhoff's theorem in Einstein-Gauss-Bonnet gravity~\cite{mn2008,birkhoff}.
In the vacuum case, Eqs. (\ref{m_v}) and (\ref{m_u}) give $m=M$, where $M$ is a constant. 
The vacuum spacetime can be completely classified by the following theorem. 
(See Proposition~1 in~\cite{mn2008} for the proof.)
\begin{The}
\label{th:vacuum}
({\it The generalized Birkhoff's theorem.})
An $n$-dimensional vacuum spacetime is isometric to one of the following:
(i) the generalized Boulware-Deser-Wheeler solution if $(D_ar)(D^ar) \ne 0$,
(ii) the Nariai-type solution if $r$ is constant, and  
(iii) the class I solution if $(D_ar)(D^ar)=k+r^2/(2\tilde \alpha)$,
where $D_a$ is a metric compatible linear connection on $(M^2, g_{ab})$.
\end{The}

The generalized Boulware-Deser-Wheeler solution~\cite{GB_BH,GBBH-solution} is given as
\begin{align}
\D s^2&=-f(r)\D t^2+f^{-1}(r)\D r^2+r^2\gamma_{ij}\D z^i\D z^j,\label{BDW1} \\
f(r) &:= k+\frac{r^2}{2\tilde{\alpha }}\left[1\mp \sqrt{1+\frac{8\kappa _n^2\tilde{\alpha }M}
{(n-2)V_{n-2}^kr^{n-1}}+4{\tilde\alpha}{\tilde\Lambda}}\right].
\end{align}

The Nariai-type solution~\cite{Lorenz-Petzold2007,md2007} is given as
\begin{align}
 \D s^2 &=-(1-\sigma \rho^2)\D t^2+\frac{\D \rho^2}
{1-\sigma \rho^2}+r_0^2\gamma _{ij}\D z^i\D z^j, \label{Nariai} \\
\sigma &:=\left[\frac{2(n-3)+2\tilde \alpha (n-5)kr_0^{-2}}{r_0^2+2\tilde \alpha k}\right]k, \label{sigma}
\end{align}
where $r_0^2$ is the real and positive root of the following algebraic equation (see~\cite{mn2008} for the existence condition): 
\begin{align}
(n-1)\tilde{\Lambda }=\frac{(n-3)k}{r_0^2}+
\frac{(n-5)\tilde \alpha k^2}{r_0^4}. \label{alg}
\end{align}
We can show $r_0^2+2\tilde \alpha k \ne 0$ since it gives a contradiction. 
The quasi-local mass of the Nariai-type spacetime is given by
\begin{align}
m=\frac{(n-2)k V_{n-2}^k r_0^{n-5}}{(n-1)\kappa_n^2}(r_0^2+2k{\tilde\alpha}),
\end{align}
where we used Eq.~(\ref{alg}) for eliminating $\Lambda$.
Thus, $m$ is non-zero for $k \ne 0$.

The class I solution~\cite{mn2008,birkhoff} exists only for $1+4\tilde \alpha \tilde \Lambda=0$ as
\begin{align}
\D s^2&=-g(r)e^{2\delta (t, r)}\D t^2 +\frac{\D r^2}{g(r)}+r^2 \gamma _{ij}\D z^i\D z^j,\label{eq:staticmetric} \\
g(r)&:=k+\frac{r^2}{2\tilde \alpha},
\end{align}
where $\delta (t, r)$ is an {\it arbitrary } function. 
The class I solution is not static in general and the quasi-local mass is zero ($m\equiv 0$).

Next, we also review the vanishing mass theorem in the asymptotically AdS spacetime for $1+4{\tilde\alpha}{\tilde\Lambda}=0$.
Equation~(\ref{qlm-sp}) shows that the quasi-local mass is non-negative for $\alpha>0$.
Then, by the combination of the asymptotic analysis and the monotonic property of $m$ on untrapped surfaces under the dominant energy condition, the following theorem is shown.
(See Proposition~7 in~\cite{mn2008} for the proof.)
\begin{The}
\label{th:positivity-sp2}
({\it Vanishing mass in asymptotically AdS spacetime with $1+4{\tilde\alpha}{\tilde\Lambda}=0$.}) 
Suppose $1+4{\tilde\alpha}{\tilde\Lambda}=0$ with $\alpha>0$ and the dominant energy condition in an $n$-dimensional asymptotically AdS spacetime.
Then, $m \equiv 0$ holds on the untrapped spacelike hypersurface.
\end{The}

In the above theorem, we employed the higher-dimensional generalization of the Henneaux-Teitelboim asymptotically AdS boundary conditions~\cite{HT1985}.
We write the metric as $g_{\mu \nu }=g^{(0)}_{\mu \nu}+h_{\mu \nu}$, where $g^{(0)}_{\mu \nu}$ is the metric of the AdS spacetime, from which deviation is represented by $h_{\mu \nu}$.
In the global coordinates $x^\mu=\{t, \rho , z^i\}$, we have
\begin{align}
g^{(0)}_{\mu \nu } \D x^\mu \D x^\nu &=-(1+\ell_{\rm eff }^{-2}\rho ^2)\D t^2+
\frac{\D \rho ^2}{(1+\ell_{\rm eff }^{-2}\rho ^2)}+\rho ^2\D \Omega _{n-2}^2, \label{AdS} \\
\ell_{\rm eff}^2&:=-\frac{1}{2\tilde \Lambda }\left(1\pm \sqrt{1+4\tilde \alpha
 \tilde \Lambda }\right), \label{ads-vac}
\end{align}
which coincide with the generalized Boulware-Deser-Wheeler solution (\ref{BDW1}) with $k=1$ and $M=0$, where $\D\Omega _{n-2}^2$ is the line element of a unit $(n-2)$-sphere.
We assume $1+4\tilde \alpha \tilde \Lambda \ge 0$ for $\ell_{\rm eff}^2$ to be real.
The fall-off conditions are
\begin{subequations}
\begin{align}
&h_{tt}={c_{tt}}\rho ^{-n+3}+O(\rho ^{-n+2}), \\ 
&h_{\rho \rho }={c_{\rho\rho}}{\rho ^{-n-1}}+O(\rho ^{-n-2}), \\
&h_{t\rho }=c_{t\rho}\rho ^{-n}+O(\rho ^{-n-1}), \\
& h_{\rho i}=c_{\rho i}\rho ^{-n}+O(\rho ^{-n-1}),\\
&h_{ti}=c_{t i}\rho ^{-n+3}+O(\rho ^{-n+2}), \\
&h_{ij}=c_{ij}\rho ^{-n+3}+O(\rho ^{-n+2}),
\end{align}
\label{bc}
\end{subequations}
where $c_{tt},...,c_{ij}$ are functions independent of $\rho $.

{\it 3. Main results:}
Now we show our main results. 
As seen before, if the spacetime is maximally symmetric or class I spacetime, the quasi-local mass $m$ is identically zero.
Indeed, for $k=1$, $\Lambda\le 0$, and $\alpha \ge 0$, its inverse also holds under the dominant energy condition. 
\begin{Prop}
\label{th:main}
({\it Vanishing mass spacetime.})
Under the dominant energy condition for $k=1$, $\Lambda\le 0$, and $\alpha \ge 0$, $m \equiv 0$ is equivalent to the maximally symmetric spacetime for $1+4{\tilde\alpha}{\tilde\Lambda} \ne 0$ and the class I spacetime (\ref{eq:staticmetric}) for $1+4{\tilde\alpha}{\tilde\Lambda}=0$.
\end{Prop}

As seen in Eq.~(\ref{ads-vac}), the special tuning between the coupling constants $1+4{\tilde\alpha}{\tilde\Lambda}=0$ allows the theory to have a unique (A)dS vacuum and become Chern-Simons gravity in five dimensions~\cite{ctz2000}.
Proposition~\ref{th:main} is proven by the combination of the following two lemmas together with Theorem~\ref{th:vacuum}.

\begin{lm}
\label{lm:untrapped}
If $m \equiv 0$ for $k=1$, $\Lambda\le 0$, and $\alpha \ge 0$, then $\theta_+\theta_-<0$, i.e., the spacetime consists of the untrapped surfaces.
\end{lm}
\noindent
{\it Proof}. 
Trivial from Eq.~(\ref{qlm2}).
\qed
\begin{lm}
\label{lm:zeromass}
Under the dominant energy condition, if $m = 0$ on the untrapped surface, then $T_{\mu\nu}= 0$.
\end{lm}
\noindent
{\it Proof}. 
By the variation formulas~(\ref{m_v}) and (\ref{m_u}), $T_{uu}=T_{vv}=T_{uv}= 0$ on the untrapped surface.
Then, the energy-momentum conservation equation $T^{a\nu}_{~~~~;\nu}=0$ gives $p r_{,a}e^{f}=0$. 
We have $r_{,a}\ne 0$ on the untrapped surface, so that $p = 0$ there, which completes the proof.
\qed

\bigskip

Then, by the combination of Proposition~\ref{th:main} and Theorem~\ref{th:positivity-sp2}, it is easy to show the following proposition about the asymptotically AdS spacetime for $1+4{\tilde\alpha}{\tilde\Lambda}=0$ with $\alpha>0$.
\begin{Prop}
\label{th:ads}
({\it Asymptotically AdS spacetime with $1+4{\tilde\alpha}{\tilde\Lambda}=0$.}) 
Suppose $1+4{\tilde\alpha}{\tilde\Lambda}=0$ with $\alpha>0$ and the dominant energy condition in an $n$-dimensional asymptotically AdS spacetime.
Then, the spacetime is represented by the class I solution (\ref{eq:staticmetric}) with $k=1$ and $\delta(t,r)$ satisfying the fall-off condition~(\ref{bc}).
\end{Prop}


{\it 4. Discussions:}
Properties of the generalized Misner-Sharp mass (\ref{qlm2}) have been fully investigated in~\cite{mn2008}.
It inherits the characteristics such as monotonicity or positivity from the Misner-Sharp mass in general relativity and is its natural counterpart in Einstein-Gauss-Bonnet gravity.
As an application, this quasi-local mass played an essential role to reveal the dynamical properties of the Gauss-Bonnet black holes~\cite{nm2008}.
Proposition~\ref{th:main} obtained in the present paper is another remarkable property in the spherically symmetric case in addition to the results in~\cite{mn2008}, which claims the equivalence between the vanishing quasi-local mass and the maximally symmetric spacetime for $\alpha \ge 0$ and $\Lambda \le 0$ with $1+4{\tilde\alpha}{\tilde\Lambda} \ne 0$.  

The case with $1+4{\tilde\alpha}{\tilde\Lambda}=0$ is exceptional in Proposition~\ref{th:main}, which admits the theory to have a unique (A)dS vacuum as well as the non-maximally symmetric vacuum solution with vanishing quasi-local mass.
Proposition~\ref{th:ads} is concerned with this exceptional case and claims that, under the dominant energy condition, even if the metric of some spherically symmetric solution of the Einstein-Gauss-Bonnet equations with $1+4{\tilde\alpha}{\tilde\Lambda}=0$ and $\alpha>0$ reduces to the AdS metric at infinity, the fall-off rate is necessarily slower than the condition~(\ref{bc}).
The generalized Boulware-Deser-Wheeler solution~(\ref{BDW1}) for $n \ge 6$ with positive $M$ is a vacuum example with such slow fall-off.
It is seen that the fall-off rate to the AdS metric changes in the case of $1+4{\tilde\alpha}{\tilde\Lambda}=0$.
This phenomenon was first pointed out in the study of the static black holes with and without the Maxwell field in the class of Lovelock gravity admitting a unique (A)dS vacuum~\cite{ctz2000}.
(The case with $k=2$ in~\cite{ctz2000} corresponds to ours.) 
Our proposition claims that it is universal in the presence of physically reasonable matter even in the highly dynamical situation.

In the $n$-dimensional Kerr-Myers-Perry-AdS spacetime, the fall-off condition~(\ref{bc}) is certainly satisfied.
Although its counterpart in Einstein-Gauss-Bonnet gravity has not been found yet, it would also exhibit the slow fall-off to the unique AdS infinity for $1+4{\tilde\alpha}{\tilde\Lambda}=0$.
Under the standard fall-off condition~(\ref{bc}), several definitions of the global mass in the asymptotically AdS spacetime have been given in Einstein-Gauss-Bonnet gravity~\cite{Padilla2003}.
However, the universal slow fall-off means that they are diverging in the case of $1+4{\tilde\alpha}{\tilde\Lambda}=0$ with $\alpha>0$.
This fact forces us to reformulate the global mass to give a finite value under the slower fall-off condition in this special case.
This problem has been investigated in Chern-Simons gravity in~\cite{cs-mass}.

Since the slow fall-off to the unique AdS vacuum has been confirmed in the vacuum case and in the presence of the Maxwell field~\cite{ctz2000}, it is naturally expected to be a universal property under the dominant energy condition also in the class of Lovelock gravity admitting a single AdS vacuum.
Our main results have been obtained essentially by using the mass variation formulas~(\ref{m_v}) and (\ref{m_u}) and the energy-momentum conservation equation.
These variation formulae are exactly the same as those in general relativity, which enable us to prove the propositions in parallel with the general relativistic case. 
In our recent paper~\cite{mn2008}, a further generalization of the Misner-Sharp quasi-local mass in general Lovelock gravity was proposed, with which the mass variation formulae were conjectured to hold.
We expect that a large part of the results obtained in the present paper and in~\cite{mn2008} is generalized in a very straightforward manner.
They will provide for us a firm ground in the research of Lovelock gravity.

\acknowledgments
The author thanks C.~Mart\'{\i}nez, M.~Nozawa, R.~Troncoso, and J.~Zanelli for comments. 
The author was supported by Fondecyt grant 1071125.
The Centro de Estudios Cient\'{\i}ficos (CECS) is funded by the Chilean
Government through the Millennium Science Initiative and the Centers of
Excellence Base Financing Program of Conicyt. CECS is also supported by a
group of private companies which at present includes Antofagasta Minerals,
Arauco, Empresas CMPC, Indura, Naviera Ultragas, and Telef\'{o}nica del Sur.



\end{document}